\documentclass[sn-vancouver]{sn-jnl}

\makeatletter \renewcommand\@biblabel[1]{#1} \makeatother

\normalbaroutside

\usepackage{caption}
\usepackage{subcaption}
\usepackage{makecell}
\usepackage{lscape}
\usepackage{siunitx}
\usepackage{scalerel}

\usepackage{glossaries}
\usepackage{booktabs}
\usepackage{colortbl}
\usepackage{xcolor}
\colorlet{tablerowcolor2}{gray!12} 
\newcommand{\rowcollight}{\rowcolor{tablerowcolor2}} %

\jyear{2021}%

\theoremstyle{thmstyleone}%

\theoremstyle{thmstyletwo}%

\theoremstyle{thmstylethree}%

\newglossaryentry{lmc}
{
    name=long-term movement consistency,
    description={long-term movement metric}
}

\newglossaryentry{smc}
{
    name=short-term movement consistency,
    description={short-term movement metric}
}

\begin{document}

\title[The impact of the COVID-19 pandemic on daily rhythms]{The impact of the COVID-19 pandemic on daily rhythms}

\author*[1]{\fnm{Nguyen} \sur{Luong}}\email{nguyen.luong@aalto.fi}

\author[2]{\fnm{Ian} \sur{Barnett}}\email{ibarnett@pennmedicine.upenn.edu}
\author[1]{\fnm{Talayeh} \sur{Aledavood}}\email{talayeh.aledavood@aalto.fi}

\affil[1]{\orgdiv{Department of Computer Science}, \orgname{Aalto Univerisity}, \city{Espoo}, \country{Finland}}

\affil[2]{\orgdiv{Department of Biostatistics, Epidemiology, and Informatics}, \orgname{University of Pennsylvania Perelman School of Medicine}, \orgaddress{ \city{Philadelphia}, \postcode{PA 19104},\country{USA}}}


\abstract{
\subsection*{Background}

The COVID-19 pandemic has significantly impacted daily activity rhythms and life routines with people adjusting to new work schedules, exercise routines, sleep patterns, and other everyday life activities.  In this work, we investigate the temporal changes of multiple daily activity rhythms and life routines throughout the course of the pandemic and highlight the disproportional changes amongst subgroups of a large cohort of working adults. Understanding the dynamics of these changes and their impacts on different groups of people is essential for creating environments where people's lives and well-being are least disturbed during such circumstances.

\subsection*{Materials and Methods}

Starting in June 2021, we conducted a year-long study to collect high-resolution data from fitness trackers as well as answers to monthly questionnaires from 128 working adults. By the end of the study, we collected over 800k data points of step count from fitness trackers and over 48k data points from answers to monthly questionnaires. Using questionnaires at different time points, we investigate how routines of exercising and working have changed throughout the pandemic for different people. In addition to that, for each person in the study, we build temporal distributions of daily step counts to quantify their daily movement rhythms. We calculate the inverse of the Earth mover's distance between different distributions of daily movement rhythms to quantify their consistency over time. Linear mixed models are used to compare the variability of daily rhythms of movement between sub-populations.

\subsection*{Results}

Throughout the pandemic, our cohort shows a shift in exercise routines, manifested in a decrease in time spent on non-walking physical exercises as opposed to the unchanged amount of time spent on walking. In terms of daily rhythms of movement, we show that migrants and those who live alone demonstrate a lower level of consistency of daily rhythms of movement compared to their counterparts. We also observe a relationship between movement and on-site work attendance, as participants who go to work (as opposed to working remotely) also tend to maintain more consistent daily rhythms of movement. Men and migrants show a faster pace in going back to work after the decrease in restriction measures that were set in place due to the pandemic. 

\subsection*{Conclusion}

Our results quantitatively demonstrate the unequal effect of the pandemic among different sub-populations. It also opens new avenues for research, to investigate why certain groups have a lower pace in going back to on-site work, the exercise level or similar daily rhythms of movement compared to prior to the pandemic, and what these mean for the future of daily life and work in the post-pandemic era. Our results can inform organizations and policymakers to provide more adequate support and adapt to the different needs of different groups in the post-pandemic era as well as during future crises.
}


\maketitle

\section{Introduction}\label{sec:intro}

The COVID-19 pandemic imposed unprecedented levels of constraints on daily lives, while disproportionately affecting life rhythms and the well-being of different subgroups of individuals \cite{chang2021mobility, suh2022disparate}. Numerous studies have identified a decline in physical activity levels \cite{bohmanStudyChangesEveryday2021, heChangesBodyWeight2021}, deterioration of mental health \cite{zandifar2020iranian, verma2022examining} and well-being \cite{zhang_unprecedented_2020, bao20202019, wang2020immediate}, changes of sleep patterns and a significant increase in digital media usage \cite{celliniChangesSleepPattern2020}. These adverse impacts have been felt to a higher degree by different groups and communities, such as women \cite{reisch2021behavioral, savolainenCOVID19AnxietyLongitudinal2021}, LGBTQ \cite{yuan2022impact, mooreDisproportionateImpactCOVID192021}, and migrant populations \cite{clark2020disproportionate, skogbergImpactCovid19Health2021}. For example, women are shown to have experienced higher levels of stress and anxiety \cite{magsonRiskProtectiveFactors2021}, more reductions in mobility \cite{caselliMobilityCOVID19Pandemic2022}, as well as a higher load in childcare \cite{pettsGenderedPandemicChildcare2021, yavorskyGenderedPandemicImplications2021} during the pandemic. Similarly, migrant populations are reported to have experienced more stress due to lower levels of social support \cite{kumarPsychologicalImpactCOVID192020}. Most of these and similar studies have investigated the overall impact and the extent to which the restriction policies have affected the lives of different subgroups. However, little is known about the impact of these policies on the daily rhythms of activities for individuals, or how people's daily rhythms and life routines have changed.

Daily rhythms of activities and their consistency over time play an important role in people's lives, both in terms of fulfilling the roles they play in society as well as at the individual level (e.g., their mental and physical well-being) \cite{carneyDailyActivitiesSleep2006, margrafSocialRhythmMental2016}. It is, therefore, crucial to understand how policies during the pandemic have led to changes in the daily rhythms of individuals. We analyze the \textit{daily rhythms of movement}, i.e., the daily movement activity patterns of people over one year during the COVID-19 pandemic. Specifically, we study how people distribute their movements throughout the course of the day, using digital records from fitness trackers combined with questionnaires answered by individuals. We explore the dynamics between daily rhythms of movement and on-site work attendance, i.e remote versus non-remote work. Furthermore, we investigate which socio-demographic factors are associated with maintaining higher levels of movement consistency and higher rates of returning to on-site work (non-remote work) throughout the course of our study.

Traditionally pen-and-paper (and later digital) surveys have been used to measure individuals' daily activity rhythms and routines \cite{monkSocialRhythmMetric1990, bondCorrelatesStructurePurpose1988}. While surveys can be an effective method to learn about the general routines of individuals, they have shortcomings in capturing daily activity rhythms and their changes, as they are prone to compliance problems and memory biases if presented at a later time. For example, in order to capture the timings of movement throughout the day, participants in the study would have to record their movements multiple times a day for more precision or just once a day if accuracy is less of a concern. This would require a high amount of effort from the respondents and cannot be feasibly used over extended periods of time. In recent years, the ubiquity of handheld and wearable devices \cite{wearableMarket2022} has gradually mitigated this compliance and memory bias issue. Personal devices leave behind a massive volume of digital traces which contain granular information about users' daily activity patterns and behaviors. Passive sensing using these devices allows capturing a person's daily rhythms in situ, which minimizes recall bias from retrospective measurements and can be used over long periods at a time due to minimal extra effort for the user. 

Before the COVID-19 pandemic, different studies have investigated daily rhythms of activity, for example, in social interactions via calls \cite{aledavoodDailyRhythmsMobile2015a}, text messages \cite{aledavoodChannelSpecificDailyPatterns2016}, and emails \cite{aledavoodDigitalDailyCycles2015}, phone usage \cite{aledavood2018social}, sleep \cite{aledavoodQuantifyingDailyRhythms2022}, or web browsing \cite{kulshrestha2021web}. These studies have shown that, despite differences between individuals, people tend to keep consistent daily rhythms and allocate a similar portion of a certain activity (e.g., making phone calls) to each section of the day (e.g., morning, evening). Another type of activity that is typically an important part of daily life is movement. Movement is a goal-directed behavior in which a specific type of movement is executed to attain a particular goal \cite{aartsHabitsKnowledgeStructures2000}. Due to the constraints that the pandemic globally imposed, most working adults' daily rhythms of movement were changed after the onset of the COVID-19 pandemic. Moreover, due to varying restrictions at different times and individual situations, people have experienced different amounts of change in their movement patterns during this time. For example, clear differences in daily rhythms of movement are observed between those who frequently commute to work and those who work remotely during the pandemic \cite{bohmanStudyChangesEveryday2021}. To the best of our knowledge, no research has yet quantified the daily rhythms of movement and their consistency over time during the pandemic.

Our study aims to examine the variations in different life routines during the pandemic and analyze the differences between these variations across socio-demographic factors, such as gender, age, and migrant status. We also aim to understand whether those routines have returned to how they were before the pandemic within this time period. Fitness tracker records and questionnaire data from 128 working adults were collected over the course of one year (June 2021-June 2022). The answers from the questionnaires were then used to show how different kinds of physical activity have changed for our study participants after the pandemic started compared to the pre-pandemic era. We proposed a metric to quantify the day-to-day daily rhythms of movement and their consistency over time, using step count data from fitness trackers. To explore the socio-demographic factors that contribute to predicting movement consistency, a linear mixed-effects model (LMM) was used \cite{raudenbush2002hierarchical}. Finally, another LMM was applied to analyze the association between this consistency metric and on-site work attendance, particularly the amount of working time spent on-site, as opposed to remote work. 

In this study, we observe a shift in exercise routines throughout the pandemic, evidenced by the significant decrease in average weekly time spent on physical exercises (other than walking) as opposed to the unchanged amount of time dedicated to walking. Secondly, we find that our participants are gradually coming back to work on-site with higher rates observed in male and migrant participants. Thirdly, our models confirm higher variability in daily rhythms of movement among migrant participants and those who live alone, as opposed to their counterparts. Fourthly, we observe a correlation between daily rhythms of movement and on-site work attendance such that individuals who prefer to work from home tend to maintain higher variability in daily rhythms of movement. In summary, our study demonstrates the gradual, disproportional changes in various life routines and daily rhythms of movement throughout the pandemic, as well as the dynamics of these changes.

\section{Materials and Methods}\label{sec:methods}

\subsection{Study design and data description} \label{sec:dataset}
In this work, we use two different datasets. The first one is the cor:ona (comparison of rhythms: old vs. new) dataset, which we collected to address the research questions in this work. The second one is the Oxford COVID-19 Government Response Tracker (OxCGRT) \cite{hale_global_2021} which is a publicly available dataset.
\subsubsection{The cor:ona study}

During June and July 2021, we recruited 128 full-time employees from a university in Finland to participate in the  cor:ona study. The main goal of this study was to examine the daily activity rhythms during different stages of the pandemic and to investigate how life routines change over time. The participants were supposed to be full-time employees of the university with no immediate plans of changing their workplace. They were asked to drop out of the study at any point when these conditions change. They could also drop out of the study for any other reason at any point in time. On average 3.2 participants left the study per month.  93 participants remained in the study for the whole year. On average the participants were part of the study for 253.5 days. Those who remained in the study for more than 6 months received a discount code from Polar which was valid for purchasing the same model or a similar fitness tracker on their website. The participants were not compensated in any other way. They were also informed that this study does not affect their employment in any way. The socio-demographic characteristics of the study participants are presented in \autoref{table:socio_demographic}. As reported in this table, we ask all participants ``Where are you from?''. They were given three choices: Finland, Europe (except Finland), or outside of Europe. Throughout this manuscript, we refer to those who indicated that they are from Finland as ``non-migrant'' and others as ``migrant''. We did not ask about the background or citizenship ofs participants. So, in this work, non-migrant and migrant categorization is solely based on the participant's answer to that question. This study was approved by the Aalto University Research Ethics Committee. 

\begin{table}[!hbt]
\sffamily
\centering
    \begin{tabular}{lr@{ }l}
    \setlength{\tabcolsep}{1pt}\\
    \toprule
    \multicolumn{1}{l}{\textbf{Attributes}} &   \multicolumn{2}{r}{\textbf{Statistics}} \\ \midrule
    
    Gender, \textit{(N\%)}& &  \\
    \hspace{3mm} Female & \hspace{2cm} 85 &  (66.4\%)  \\
    \hspace{3mm} Male  & \hspace{2cm} 42 & (32.8\%)     \\
    \hspace{3mm} Non-binary & \hspace{2cm} 1 & (0.8\%) \\
    
    Role at university, \textit{(N\%)} & &   \\
    \hspace{3mm} Academic staff & \hspace{2cm} 56 & (43.8\%)     \\
    \hspace{3mm} Service staff  & \hspace{2cm} 72 & (56.2\%)     \\
	
    Age, \textit{(N\%)} & & \\
    \hspace{3mm} 25-35 & \hspace{2cm} 60 & (46.5\%)    \\
    \hspace{3mm} 36-50 & \hspace{2cm} 47 & (37.0\%)     \\
    \hspace{3mm} 51-66 & \hspace{2cm} 21 & (16.5\%)    \\
    
    Origin, \textit{(N\%)}   \\
    \hspace{3mm} Finland               & \hspace{2cm} 91 & (70.9\%)     \\
    \hspace{3mm} Europe except Finland & \hspace{2cm} 15 & (11.8\%)     \\
    \hspace{3mm} Outside of Europe     & \hspace{2cm} 22 & (17.3\%)  \\

    \bottomrule
    \end{tabular}
\caption{Socio-demographic characteristics of the cor:ona study participants.}
\label{table:socio_demographic}
\end{table}

\subsubsection*{Questionnaires}

The participants completed a baseline questionnaire upon entering the study. This questionnaire included questions about socio-demographic information and life routines (see \autoref{table:routine_questions}). All participants filled out the baseline questionnaire during late June and early July of 2021, approximately 15 months after the start of the pandemic (mid-March 2020). They were  asked to compare their activities in those 15 months (referred to in our study as the ``early stage of the pandemic'') with the 15 months prior to the pandemic (referred to in our study as ``prior to the pandemic''). The participants received a short version of the baseline questionnaire every month. 
On average 103.5 participants answered the questionnaires each month. The participants completed an exit survey at the end of the study to assess their life routines during the year-long study period since they filled out the baseline questionnaire (referred to as the ``late stage of the pandemic'' in this work).

\subsubsection*{Wearables}

\begin{figure}[!ht]
    \centering
    \includegraphics[width=\textwidth]{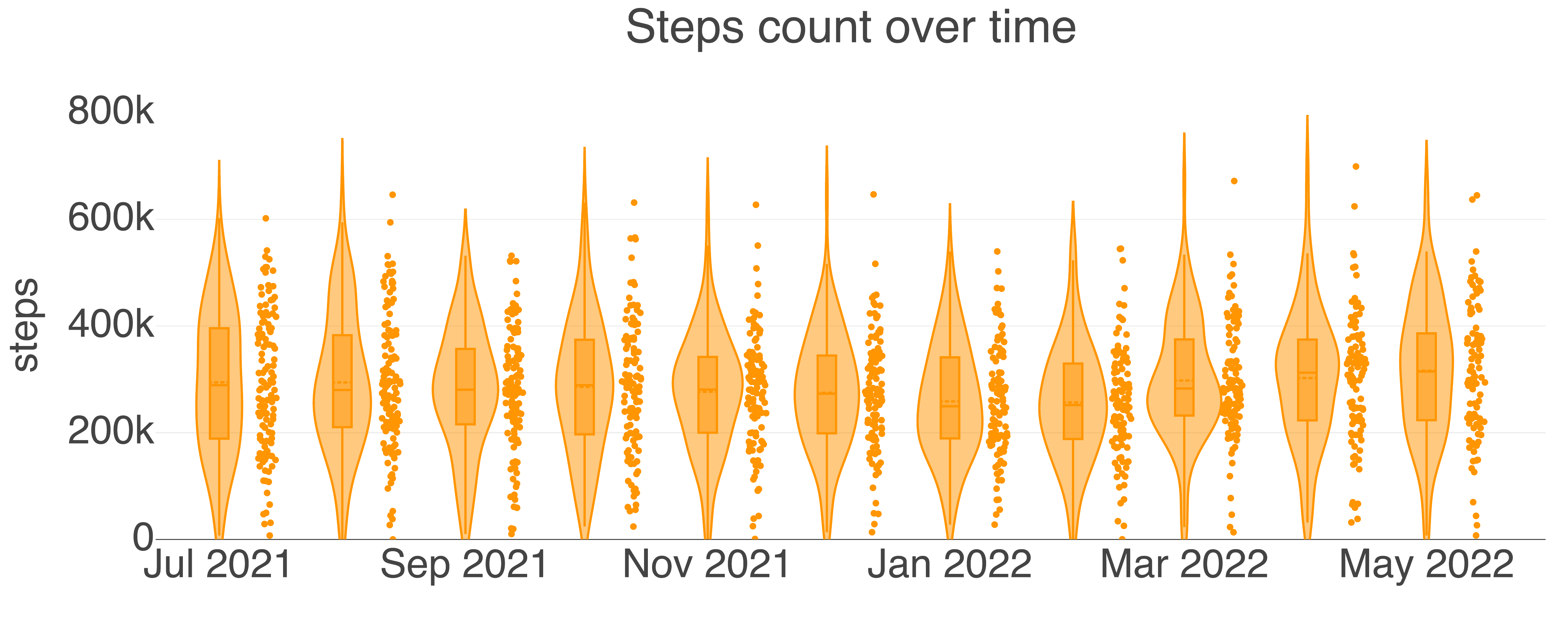}
     \caption{Step count values of all participants over the course of the study.}
     \label{fig:step-count-demo}
\end{figure}

Participants were loaned new Polar Ignite fitness trackers (N=121). 7 participants joined the study using their personal Polar fitness trackers. According to the manufacturer, the Ignite fitness trackers have 5 days of battery operating time in normal mode and up to 17 hours in training mode with GPS on \cite{polarIgniteBattery2022}. With users' consent, we collected their personal data from Polar's platform using Polar Accesslink API \cite{polarIgniteApi2022}. This includes the step counts for each person in the study with hourly time resolution. In total, our data set consists of 810,111 data points of step count and 48,228 data points from survey answers. Fig. \ref{fig:step-count-demo} shows the monthly step count of all participants over the course of the study. 

\subsubsection{Oxford COVID-19 Government Response Tracker} \label{sec:stringency_index}

The Oxford COVID-19 Government Response Tracker (OxCGRT) is a project designed to capture government policies related to closures and containments in more than 180 countries \cite{hale_global_2021}. The project evaluates national and sub-national restriction policies and provides a list of composite indices measuring their extent. The sum of these indices is called the stringency index, which is a number between 0 and 100 (higher values indicating stricter policies) and is provided for each day during the pandemic for each region. In this work, we use the stringency index for Finland which evaluates the containment and closure policies of the country throughout our study. We use these data to assess the connection between the policies and movement/on-site work attendance. 

\subsection{Measuring time allocation across activities}

To quantify the changes in exercise and work routines during different stages of the pandemic, relevant questions were asked in the baseline and exit questionnaires (related to the amount of walking and non-walk exercises), as well as on a monthly basis (related to the percentage of working time spent working on-site in the past month). For instance, questions about non-walking activities, such as ``Prior to the pandemic; how many hours per week on average did you walk'' and ``During the pandemic; how many hours per week on average have you walked?'' were asked in the baseline questionnaire while the question ``In the late stage of the pandemic; how many hours per week on average have you walked?'' was asked in the exit questionnaire. The detailed questions are presented in \autoref{table:routine_questions}.

Given that the activities being studied were not normally distributed, the analysis of the time allocation for each stage of the pandemic and the comparisons between different populations were performed using non-parametric statistical tests. Specifically, the Wilcoxon-signed ranked test was used to compare time allocation for each stage of the pandemic, and the Mann-Whitney U test was employed to compare different populations.

\subsection{Quantifying daily rhythms of movement and their consistency over time}

\subsubsection*{Daily rhythms of movement}\label{sec:step_distribution}

\captionsetup[subfigure]{labelformat=empty, position=top, singlelinecheck=off, justification=raggedright}

\begin{figure}[!hbt]
    \centering
    \includegraphics[width=\linewidth]{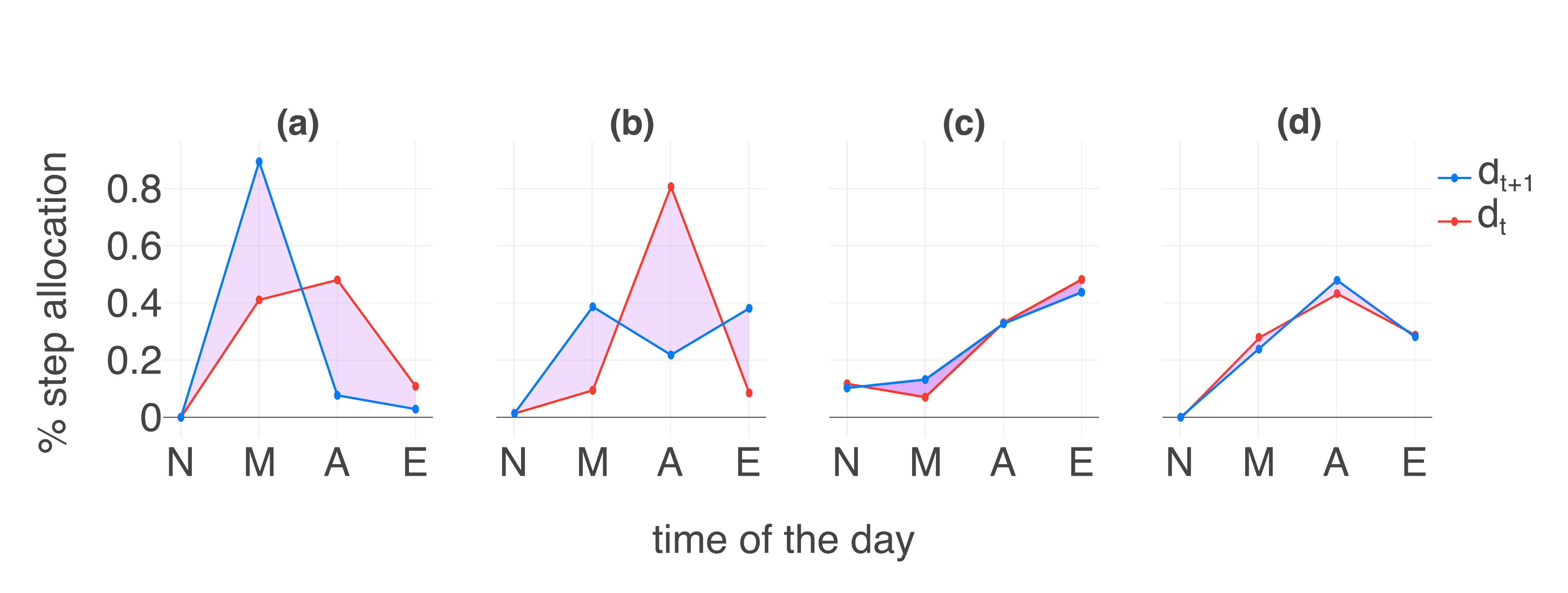}
    \caption{Visualization of movement consistency. A sample of 4 individual-level daily step allocations to demonstrate the level of movement consistency over time. The red line illustrates the step count distribution of ${\bf d}_t$ and the blue line illustrates the distribution of  $d_{t+1}$. The markers indicate the distribution of steps in a single time segment (\textbf{N}: night midnight-6am, \textbf{M}: morning 6am-noon, \textbf{A}: afternoon noon-6pm, \textbf{E}: evening 6pm-midnight). The purple area indicates the distribution dissimilarity between two consecutive days. In panels \textbf{(a)} and \textbf{(b)}, this area is considerably larger than that of panels \textbf{(c)} and \textbf{(d)}, highlighting higher variability in day-to-day rhythms of movement.
    }
    \label{fig:consistency_lh}
\end{figure}

To quantify the daily rhythms of movement of people, we calculate the temporal distribution of the daily step counts for each person in the study. First, we aggregate step counts at hourly resolution into larger time segments. We compute the number of steps for each person and each day in the four different time segments: midnight-6am (night), 6am-noon (morning), noon-6pm (afternoon), and 6pm-midnight (evening). We then normalize the step counts by dividing them by the total number of steps on that day. This enables us to assess the proportion of steps taken by the individual during each time segment on a given day. 

Due to the inherent differences in movement rhythms between workdays and weekends, we consider them separately. In this work, Saturday and Sunday, are referred to as ``weekends'' and other days as ``workdays''. We denote the distribution of step counts of the day $t$ as ${\bf d}_t$, which represents the daily rhythms of movement. The choice of four as the number of time segments in a day and their timing is based on prior work in quantifying activity rhythms \cite{aledavoodDailyRhythmsMobile2015a}. We also evaluated the impact of different time-segmenting strategies on the consistency metrics (see \autoref{secA4}). The results were compared across various segmenting strategies and we found a consistent pattern in all cases, indicating that our method is robust to the granularity of time segmentation.

\subsubsection*{Short-term movement consistency} \label{sec:routine_quantification}

\begin{figure}[!ht]
     \centering
     \includegraphics[width=\textwidth]{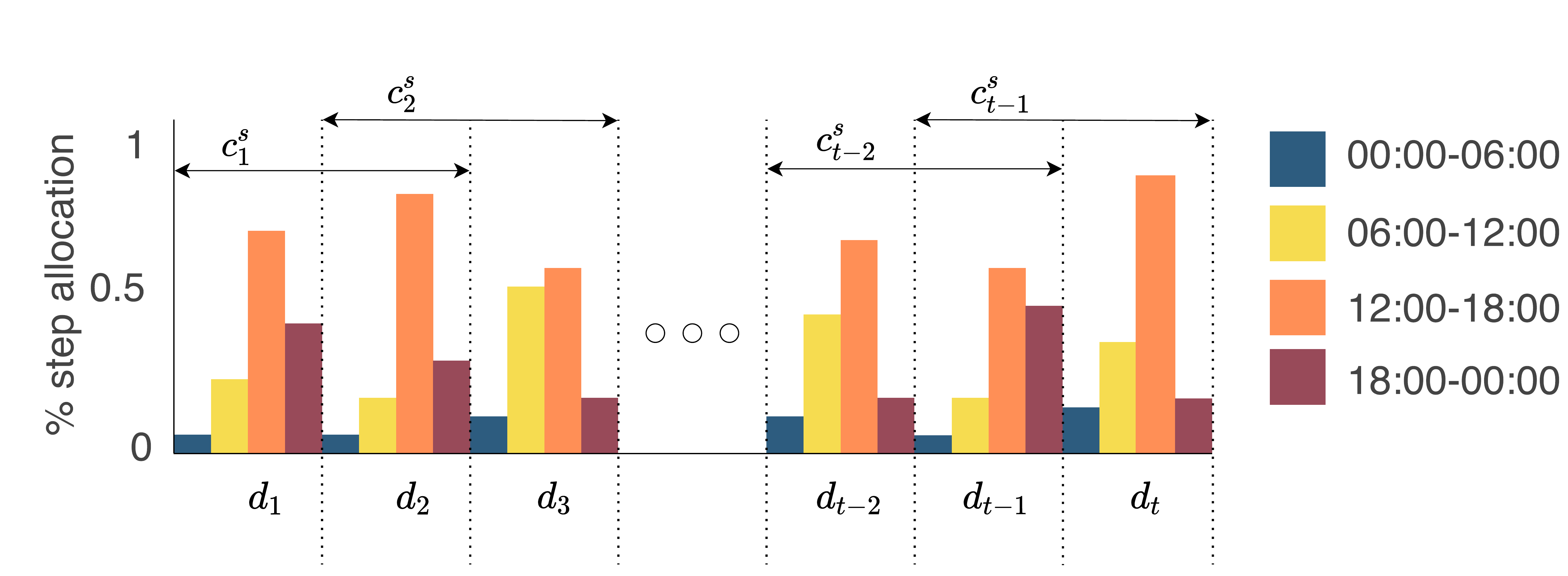}
     \caption{Short-term movement consistency computation. The short-term movement consistency is denoted as $c^s_t=1/D({\bf d}_t,{\bf d}_{t+1})$ and quantified as the inverse of the distance in step count distribution between ${\bf d}_t$ and  ${\bf d}_{t+1}$ of an individual.} 
     \label{fig:short-parc}
\end{figure}

The daily rhythms of movement of a person can vary both in the short and long term. We propose a metric called \textit{short-term movement consistency} ($c^s$) to measure how the daily rhythms of movement of a person vary from one day to the next. We quantify the movement consistency as the inverse of the Earth mover's distance (EMD) \cite{rubner_earth_2000} of the distributions of daily rhythms of movement between two consecutive days: 

$$
c^s_t =  {1}/{D({\bf d}_t, {\bf d}_{t+1})}\,.
$$
The smaller the distance between each pair of daily rhythms of movement is, the more consistent the daily rhythms have been. Fig. \ref{fig:short-parc} depicts how short-term movement consistency is calculated.

To assess the changes in movement consistency over time, we calculate the average short-term movement consistency for each participant for each calendar month.
For each month, the average short-term metric is defined as
$$
\bar{c}^s_{\mathrm{wd}} = \sum_{t \in \hat{N}_{\mathrm{wd}} }c^s_t/ \lvert \hat{N}_{\mathrm{wd}} \rvert \,,
$$
where $\hat{N}_{\mathrm{wd}}$ is the set of workdays within that month excluding Fridays. For weekends the average is defined as
$$
\bar{c}^s_{\mathrm{we}} = \sum_{t \in \hat{N}_{\mathrm{we}}}c^s_t/\lvert \hat{N}_{\mathrm{we}} \rvert\,, 
$$
where $\hat{N}_{\mathrm{we}}$ is the set of Saturdays within that month. So in this case, each $c^s_t$ is calculated by comparing the daily rhythm of a Saturday to the day after (Sunday).

\subsubsection*{Monthly and long-term movement consistency} \label{sec:long_routine_quantification}

\begin{figure}[!ht]
    \includegraphics[width=\textwidth]{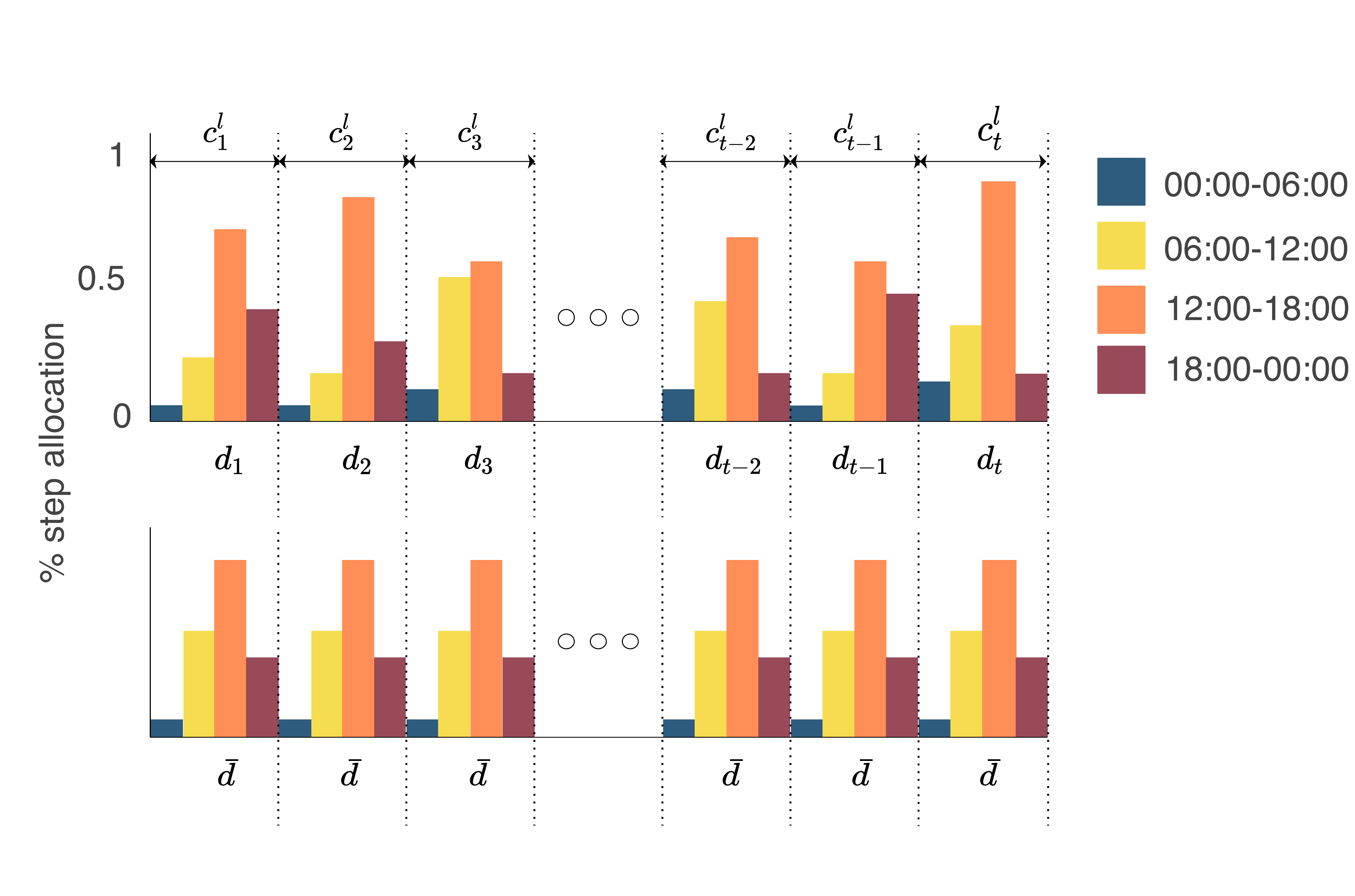}
    \caption{\Gls*{lmc} calculation. The long-term movement consistency is denoted as $c^l_t=1/D({\bf d}_t,\bar{{\bf d}}^l)$ and quantified as the inverse of the distance in step count distribution between ${\bf d}_t$ and the average distribution $\bar{{ \bf d}}^l$ of an individual.}
    \label{fig:long-parc}
\end{figure}   

\Gls*{smc} is sensitive to changes in daily rhythms that happen on one day and make the person's daily rhythm different compared to the day before. However, if the daily rhythm has only minor changes from one day to the next but follows a certain trend on a longer time scale, the short-term movement consistency is not suitable to capture this. In longer timescales, the rhythm consistency can change to adapt to circumstances that last longer such as changes in seasons or changes in one's work or family life. To capture these kinds of changes, we propose two metrics: \textit{monthly movement consistency} and \textit{long-term movement consistency}. These metrics measure the difference between an individual's one-day movement rhythm and their average (baseline) rhythm. The baseline which is denoted as $\bar{{\bf d}}$, can be calculated based on the average movement daily rhythms of a person over an extended period of time. We denote the monthly baseline as $\bar{{\bf d}}^m$ and the long-term baseline, which consists of the whole time a person has participated in the study, as $\bar{{\bf d}}^l$.  We define monthly movement consistency, $c^m_t$, as the inverse of the distance between the daily rhythms of movement ${\bf d}_t$ and the monthly baseline $\bar{{\bf d}}^m$,
 $$c^m_t=1/D({\bf d}_t,\bar{{\bf d}}^m)\,.$$
Similarly, we define the long-term movement consistency as 
$$c^l_t=1/D({\bf d}_t,\bar{{\bf d}}^l)\,.$$
 Fig. \ref{fig:long-parc} depicts a schematic visualization of the long-term movement consistency.
 
 Both these calculations will lead to one value for movement consistency for each day and each participant. Similar to short-term movement consistency we compute means of these values for weekdays and weekends separately, i.e., we define
 \begin{eqnarray}
\bar{c}^m_{\mathrm{wd}} = \sum_{t \in N_{\mathrm{wd}} }c^m_t/ \lvert N_{\mathrm{wd}} \rvert \,, \nonumber \\
\bar{c}^m_{\mathrm{we}} = \sum_{t \in N_{\mathrm{we}}}c^m_t/\lvert N_{\mathrm{we}} \rvert\,, \nonumber
 \end{eqnarray}
where $N_{\mathrm{wd}}$ is the set of workdays and $N_{\mathrm{we}}$ is the set of days in weekends within a month. $\bar{c}^l_{\mathrm{wd}}$ and $\bar{c}^l_{\mathrm{we}}$ are calculated similarly to $\bar{c}^m_{\mathrm{wd}}$ and $\bar{c}^m_{\mathrm{we}}$, by replacing $c^m_t$ with $c^l_t$ in the two formulas above respectively.

\subsection{Models: Notations and selection}\label{sec:model_creation}

We propose three models to evaluate the relationship between movement
consistency and socio-demographic factors. We consider three different
consistency metrics as the dependent variable (DV): short-term weekday
(Model 1a), long-term weekday (Model 1b), and long-term weekend (Model
2). We do not consider short-term weekend metric since it is just the difference between Saturday and Sunday and long-term data is needed for this comparison.
We take into account the following independent variables (IVs): age,
gender, origin, cohabitation status (whether the subject is living with someone
else), children status (whether the subject has children), and role at the
university (academic staff or service staff). Additionally, we consider a two-way interaction between gender and children status to account for the role of
gender in childcare duty.

We build another model (Model 3) to evaluate the relationship between on-site work attendance (the fraction of time an individual works on-site each month) and movement consistency. This variable is the percentage of working time spent on-site in the past calendar month and is reported in the monthly surveys. To analyze the association between movement consistency and on-site work attendance, in addition to the above IVs, we also include the average long-term workday consistency metric which is calculated by getting the mean of all workday consistencies within the month. All the above response variables are measured at monthly intervals. For Models 1a, 1b, and 2, we use the monthly average of the short and long-term movement consistencies respectively for each month. 

To control for repeated measurements, we employ a linear mixed effects model (LMM) \cite{raudenbush2002hierarchical}. For Model 1a, 1b, and 2, we include participants as random effects to account for the intra-individual difference. For Model 3, we incorporate calendar months (July, August, September, etc.) as random effects to account for the variations in on-site work rates due to restrictions policies. We do not consider individual-specific random effects for Model 3 due to insufficient data to incorporate these two levels of random effects. Instead, we capture the individual-level variation with extensive fixed effect covariates. For individual $i$ at calendar month $j \in \{7 (July), 8 (August),..., 5 (May)\}$, we denote $Y_{ij}$ as the variable of interest and $x_{ij}$ as the covariate. The intercept for the random effect is denoted as $u_j$. We formally describe the models as follows:\\

Model 1a, 1b, 2: $Y_{ij} = \beta_0 + \beta_1x_{ij1} + \beta_2x_{ij2} + .. + \beta_7x_{ij7} + u_j + \epsilon_{ij} $,

Model 3: $Y_{ij} = \beta_0 + \beta_1x_{ij1} + \beta_2x_{ij2} + .. + \beta_7x_{ij7} + \beta_8x_{ij8} + u_j + \epsilon_{ij}$,\\

where the IVs of individual $i$ during month $j$ are:\\

$x_{ij1}$:=age group, $x_{ij2}$:=gender, $x_{ij3}$:=origin,

$x_{ij4}$:=live with someone, $x_{ij5}$:=have children, 

$x_{ij6}$:=gender $\times$ live with someone, $x_{ij7}$:=gender $\times$ have children,

$x_{ij8}$:=monthly average of long-term workday movement consistency.\\

We report estimates of coefficients. The \textit{p}-values and 95\% confidence intervals for the estimates of coefficients are calculated using a parametric bootstrap. We perform variance inflation factor (VIF) analysis for all models. The max GVIF for any predictor is 2.2 (gender $\times$ have children), which is under the recommended threshold \cite{obrienCautionRegardingRules2007}.

\subsubsection*{Data exclusion and pre-processing}

We exclude the participants with less than 20\% available step count data points during the one year of the study. This results in a total of 111 participants. For days with no step count recorded, we consider them as missing data and exclude them from the analysis. To account for the missing data due to participants dropping out or not wearing the devices frequently enough, for each month, we exclude participants if they do not have step count data for 5 workdays or 2 full weekends. For gender-related analysis, we exclude non-binary participants to preserve their privacy. Since people tend to have more variability in daily rhythms while on holiday and thus distorting the relationship between movement consistency and on-site work attendance, we exclude the participants with more than 7 days of leave in a given month for on-site work attendance analysis (Model 3). The number of days on leave for each participant is asked monthly through the questionnaires. We standardize all numerical variables in all models.

\section{Results}\label{sec:results} 

\subsection*{Changes in daily activities after the onset of the pandemic} \label{sec:results-routines-changes}

\begin{table}[!hb]
\sffamily
\centering
    \begin{tabular}{lrr@{}lr@{}l}
    \setlength{\tabcolsep}{1pt}\\
     \multicolumn{1}{c}{\textbf{Attributes}} & \multicolumn{5}{c}{\textbf{Pandemic stage}} \\ 
     \toprule
     & \textbf{Pre} & \textbf{Early} & & \multicolumn{2}{l}{\textbf{Late}} \\\midrule
     
     \rowcollight Avg. hours per week for walking & & & & & \\

    \hspace{1cm} All     & 5.33   & 4.83 & & 5.97 &   \\
     
     \hspace{1cm} Male     & 4.53   & 3.67 & & 4.93 &   \\
     \hspace{1cm} Female    &  5.68 & 5.43  & & 6.61 &    \\
     \hspace{1cm} non-migrant    &   5.37  & 5.17 & & 6.07 &    \\
     \hspace{1cm} migrant   &  5.33  & 4.18  & & 6.00 &     \\
    \rowcollight Avg. hours per week for non-walking exercises & & & & & \\
     \hspace{1cm} All  &  3.62 & 2.58 & ***  & 2.76 & ** \\
     \hspace{1cm} Male  &  3.71 & 3.37 &   & 3.04 &  \\
     \hspace{1cm} Female &  3.54 & 2.21 & ***  & 2.57 & ** \\
     \hspace{1cm} non-migrant & 3.42 & 2.15 & ***  & 2.45 & **  \\
     \hspace{1cm} migrant  & 4.03 &  3.51 & & 3.33 &  \\ 
     \rowcollight \% of working time spent on-site  & & & & &       \\
     \hspace{1cm} All      &    83.64   & 9.94 & *** & 32.56 & *** \\
     \hspace{1cm} Male  & 83.42 & 13.57 & *** & 43.27 & *** \\
     \hspace{1cm} Female   & 83.75  & 8.20 & *** & 27.44 & *** \\
     \hspace{1cm} non-migrant   &  82.38  & 8.62 & *** & 27.87 & *** \\
     \hspace{1cm} migrant  & 86.54  & 12.97 & *** & 43.32 & *** \\
     \bottomrule
    \end{tabular}
\caption{Average amounts of activities at different stages of the pandemic compared to the pre-pandemic time. The average amount of weekly walking and non-walking exercise, as well as the percentage of working time spent on-site, is represented for three time periods: pre-pandemic, early, and late stages of the pandemic. The comparison of the percentage of work time spent on-site during the late stage of the pandemic is based on participants who have responded to at least 5 monthly surveys. The pre-pandemic stage is set as the reference level and the mean of each of the other two stages is compared to the pre-pandemic value of the mean for a given activity and sub-population. All comparisons are made using Wilcoxon signed rank test. Asterisks denote the significance of the results. *$p<0.05$, **$p<0.01$, ***$p<0.001$.}
\label{table:routine_estimates}
\end{table}

The shift from on-site to remote work has been extensively documented since the outbreak of the COVID-19 pandemic, as mandated lockdowns were implemented to curb the spread of the virus. Additionally, with public indoor spaces such as gyms being restricted, a shift to other forms of exercise became necessary and even encouraged. For example, at the university where our study was conducted, employees were often recommended to have their remote meetings while taking (individual) walks outdoors. To determine how these circumstances changed daily activities, we examine the amount of remote work vs. on-site work, and walking and non-walking exercises over time. For this purpose, we use the reported values by the study participants in the questionnaires
(see \autoref{table:routine_questions} for a detailed list of questions that participants were asked).

In \autoref{table:routine_estimates}, we show the reported estimate of time allocation for on-site work and exercises in three periods: pre-, early, and late stages of the pandemic. The average hours for walking weekly do not differ significantly between the pre-pandemic stage against the early and late stages. 
The amount of hours spent on non-walking exercises drops significantly when the early stage (mean = 2.58 hours, SD = 2.51 hours) and the late stage (mean = 2.71, SD = 2.38 hours) average weekly activity hours are compared to the pre-pandemic level (mean = 3.60 hours, SD = 2.75 hours). On the sub-population level, the decrease in non-walking exercises is significant for female and non-migrant participants but not for male and migrant participants. This suggests that the former two groups have not been as successful as the latter in maintaining their non-walking exercises after the pandemic and the restrictions started. The monthly percentage of time spent on-site drops significantly from the pre-pandemic (mean = 83.64$\%$, SD = 22.67$\%$) to the early stage of the pandemic (mean = 9.94$\%$, SD = 19.39$\%$) and slowly recovers in the late stage (mean = 32.56$\%$, SD = 25.16$\%$). 

\begin{figure}[!htbp]
     \centering
     \includegraphics[width=\linewidth]{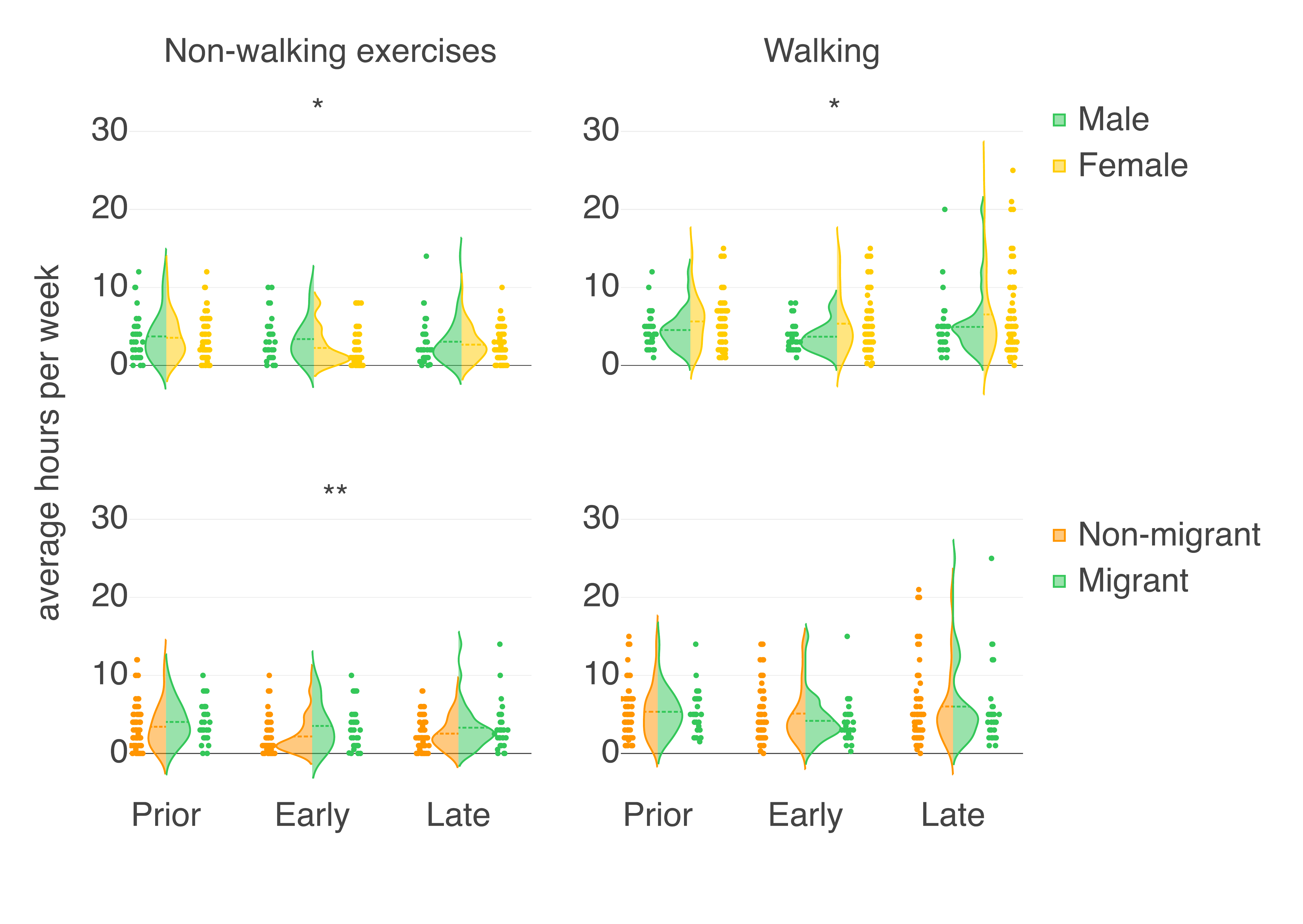}
     \caption{Time allocation for different activities of sub-populations during each stage of the pandemic. Comparisons are made using Mann-Whitney U test. Asterisks denote the significance of the results. *$p<0.05$, **$p<0.01$, ***$p<0.001$.} 
     \label{fig:activities_comparison}
\end{figure}

Distributions of non-walking and walking exercises and sub-population comparisons are shown in  Fig. \ref{fig:activities_comparison}. There is no significant difference in the time spent on walking between male and female sub-populations during the pre-pandemic stage ($U= 790.0, p=0.06$). However, during the early stage of the pandemic, females show a significantly higher amount of time spent walking ($U=727.5, p=0.02$). Conversely, in regards to non-walking exercises, females tend to spend significantly less time during the early stage of the pandemic compared to males ($U=709.5, p=0.01$).

Compared to the early stage, during the late stage of the pandemic, there is an increase in the trend for walking ($U=1333.0, p=0.02$) and on-site work attendance ($U=168.0, p<0.001$) as individuals gradually return to pre-pandemic levels. On the other hand, the trend for non-walking exercises stay relatively the same ($U=1210.0, p=0.14$), suggesting that walking may be more consistently incorporated into people's daily routines to compensate for other forms of exercise.

\subsection*{Movement consistency varies among different sub-populations.}

\begin{figure}[!htp]
     \centering
     \includegraphics[width=\textwidth]{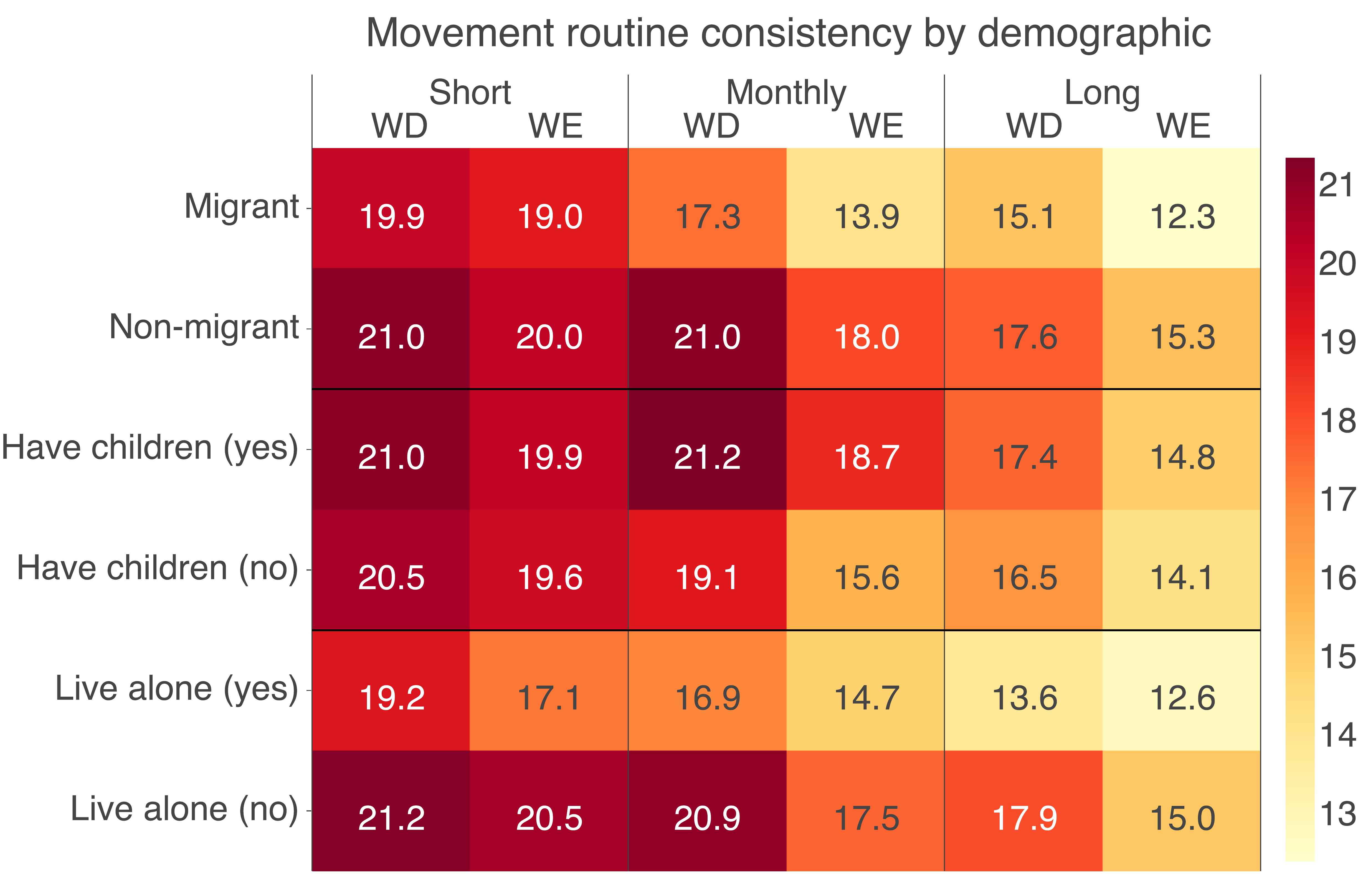}
         \caption{Movement consistency variability by socio-demographic factors. Lower values indicate lower consistency. Weekday is denoted as WD and weekend is denoted as WE. The monthly consistency is measured as the inverse of the distance between the one-day distribution against the monthly average distribution. For all groups, weekend variability is generally higher than variability of workdays. Migrant participants tend to have the lowest consistency level.} 
     \label{fig:consistency_demographic}
\end{figure}

Individuals may have different variability in daily rhythms of movement due to societal responsibilities, such as parenting, or the nature of their occupation. Fig. \ref{fig:consistency_demographic} shows the variations in movement consistency among sub-populations, with a greater contrast observed in long-term daily rhythms of movement. We formally examine these differences by comparing the daily rhythms of movement of all participants while controlling for socio-demographic factors. We present the results of Model 1a and 1b (short-term and long-term weekend movement consistency) in \autoref{table:weekday-long-consistency}. Both models suggest that participants living alone tend to have 
lower short-term ($\beta=-0.33, p=0.022$) and long-term ($\beta=-0.78, p<0.001$) movement consistency. Similarly, migrant participants have 
lower long-term movement consistency ($\beta=-0.61, p=0.004$) but the difference in short-term movement consistency is not significant
($\beta=-0.19, p=0.210$). Age, role at university, gender, and children's status are not significant predictors for short-term and long-term movement consistency. The result for Model 2 (long-term weekend consistency) is presented in \autoref{secA2}.

\begin{table}[!htp] 
 \centering 
 \sffamily 
 \footnotesize 
 \begin{tabular}{lrr@{}lrr@{}l} 
 \toprule & 
 \multicolumn{3}{c}{Model 1a (short-term)} & 
 \multicolumn{3}{c}{Model 1b (long-term)} \\ 
 \cmidrule(l){2-6} & 
 Est.      & 
 95\% CI  & 
 & 
 Est.      & 
 95\% CI\\ 
 \midrule 
 
 (Intercept) & 
 0.22 &	
 -0.08 – 0.53 &  
 & 
 \textbf{0.48} &	
 \textbf{0.06 – 0.91} & 
 *      \\ 

Role (service staff) & 
 -0.08 &	
 -0.36 – 0.21 & 
 & 
 -0.17 &	
 -0.56 – 0.22 \\ 
 Gender (male) & 
-0.16 &	
-0.45 – 0.15 & 
& 
-0.11 &	
-0.30 – 0.08 \\ 

Live alone (yes) & 
\textbf{-0.33} &
\textbf{-0.61 – -0.05} & 
*& 
\textbf{-0.78} &
\textbf{-1.15 – -0.40} & 
*** \\ 

Have children (yes) & 
-0.10 &
-0.40 – 0.19 & 
& 
-0.29 &	
-0.70 – 0.11 
& \\ 
  
Origin (migrant)  & 
-0.19	& 
-0.50 – 0.10 &  
& 
\textbf{-0.61} & 
\textbf{-1.01 – -0.20} & 
** \\ 
 
Age & 
0.01 & 
-0.12 – 0.14 &
& 
-0.02 &	
-0.19 – 0.15 &   
\\ 
 
Have children (yes) x Gender (male) & 
0.22 &	
-0.26 – 0.70  & 
& 
0.31 &	
-0.37 – 0.98   \\ 
 
 \midrule
 \textbf{Random effects} \\
 $\sigma^2$ & 0.74 & & & 0.34\\
 $\tau_{00_{participant}}$ & 0.25 & & & 0.61\\
 ICC & 0.25 & & & 0.64\\
 $N _{participant}$ & 111 & & & 111\\
 Observations & 885 & & & 912\\
 Marginal $R^2$ / Conditional $R^2$ & 0.025 / 0.271 & & & 0.131 / 0.687\\
 \bottomrule 
 \end{tabular} 
 \caption{Results for socio-demographic variables predicting short and long-term workday consistency. Asterisks denote the significance of the results. *$p<0.05$, **$p<0.01$, ***$p<0.001$.} 
 \label{table:weekday-long-consistency} 
 \end{table}

\subsection*{Movement consistency positively correlates with working on-site}

We showed that the rate of returning to on-site work varies among sub-populations. On-site work attendance may also be linked to movement consistency as individuals who go to work tend to have a more structured schedule and therefore exhibit a more consistent daily rhythm of movement. Thus, we would expect to see a correlation between the rate of on-site work and movement consistency. Model 3 confirms a positive correlation between long-term movement consistency and on-site work attendance, as one unit increase in movement consistency results in a 7\%$\pm$2\% (95\% CI) increase in time spent on-site. Among all demographic factors, male and migrant participants tend to spend 12\%$\pm$6\% and 13\%$\pm$5\% (95\% CI) more time on-site on average than their counterparts respectively. Individuals who live alone tend to spend more time at work on-site, with an increase of 13\%$\pm$5\% (95\% CI) compared to those who live with others. The full results are displayed in \autoref{secA2}.

The relationship between movement and the percentage of on-site work can also be influenced by restriction policies. We hypothesize that the restriction on public places leads to changes in daily rhythms of movement, with more consistent rhythms observed during periods of stricter restriction as a result of limited socializing opportunities. To test this hypothesis, we interpret the random intercepts from Model 3 to analyze the variability at each random effect level. From Model 3, random intercepts measure the inclination to work on-site each month. When comparing this property against the stringency index (section \ref{sec:stringency_index}), we observe a positive relationship: participants are willing to work from home when the policy is tightened ($r=-0.897, p < 0.001$) (see Figure \ref{fig:si_vs_re}). Similarly, we retest Models 1a and 1b but used the calendar month as a random effect. The random intercepts from Model 1b measure the variability of long-term movement consistency each month. The results show that this variability negatively correlates with the stringency index ($r=-0.836, p=0.001$), meaning that stricter restrictions are linked to less consistent daily movement rhythms. Taken together, the results suggest that the participants generally adhered to the restrictions and had a more diverse range of activities while working remotely.

\section{Discussion}\label{sec:discussion}

Our study presents a high-resolution and long-term data set of working adults during the COVID-19 pandemic. We focus on four areas of investigation: (1) comparing time allocation for different daily activities between pre-, early, and late pandemic stages; (2) introducing new methods to infer daily movement rhythms and their consistency from fitness tracker data; (3) examining variability in daily rhythms of movement among different sub-populations; and (4) evaluating the connection between daily rhythms of movement and on-site work attendances. We propose a method using high-resolution step count data to infer day-to-day and long-term differences in movement rhythms. By controlling for factors such as age, gender, and role at the workplace, higher variability in long-term movement consistency among migrants and those living alone was observed. Additionally, long-term movement consistency and the likeliness to work on-site were positively linked. 

The amount of walking remained unchanged during the transition from pre- to the early pandemic stage and increased during the transition from the early to late stage. These changes could be attributed to a variety of factors. In the early pandemic stage, walking was a common leisure activity \cite{gupta2021cross} and an effective coping mechanism against boredom \cite{lohiniva2021learning}. During the pandemic, travel behaviors changed significantly as private transport modes (walking, bicycle, motorcycle, car) were preferred over public transport modes (bus, tram, train) due to perceived risks of COVID-19 transmission \cite{barbieri2021impact}. Moreover, since there was no longer a need to commute to work, commuting distances were shorter, and the majority of trips were made for shopping purposes \cite{abdullah2020exploring}, which encouraged walking as a mode of transportation even more.

For all participants, the amount of time for non-walking exercises dropped significantly in the early stage and did not increase during the transition from the early to late stage. Gender-wise, the decline was significant among females but not in males. This result contradicts prior studies that have shown a larger decrease in physical activity among males during the early stage of the pandemic, which was primarily attributed to differences in exercise habits and intensity between genders \cite{castaneda2020physical, giustino2020physical, maugeri2020impact}. Those studies argued that men tend to engage in exercise more frequently, so a sudden reduction in the amount of physical activity is more noticeable. Nonetheless, the difference in non-walking exercises between the genders became insignificant during the late stage.

Daily movement rhythms were consistent within the short term for individuals, with little difference found between sub-populations. However, higher variability was found in monthly (the metric between short and long-term) and long-term measurements for migrants and those living alone. Daily changes in movement rhythms may indicate an effort to increase social interaction as a way of combating loneliness \cite{rokach1998coping}. Results from Model 3 showed that the same sub-populations had a higher rate of working on-site, suggesting that they may be using going to work as a means of alleviating social isolation. During the March 2022 survey, a question was presented to the participants regarding their reasons for returning to on-site work. The findings revealed that 53\% of migrant participants cited social interaction as their main motivation, whereas only 35\% of non-migrant participants did so. This difference highlights the significance of social interaction for individuals experiencing loneliness, who may be more impacted by feelings of social isolation compared to others.

Males had a higher on-site work attendance than females. Despite numerous studies addressing a higher load of childcare duties among females during the COVID-19 pandemic \cite{reisch2021behavioral, savolainenCOVID19AnxietyLongitudinal2021, ramos2020women, viglione2020women}, we did not find a significant effect of gender-specific childcare duties on the rate of on-site work attendance. It is possible that, within our cohort, childcare duties were spread evenly thus mitigating the gender bias. Additionally, the absence of a complete lockdown in Finland during the school periods and the reopening of schools could have played a role in reducing gender disparity, as school openings have been linked to a decrease in gender-bias mobility \cite{reisch2021behavioral}. Nonetheless, other factors could contribute to this gender difference. One possible explanation could be the tendency for men to have a higher risk-taking behavior \cite{gustafsod1998gender, byrnes1999gender}, which enables them to disregard the fear of COVID as they returned to on-site work.
During the pandemic, this disparity in risk perception between males and females has also persisted. For instance, one study discovered a correlation between being male and a lower propensity to engage in health-protective behaviors \cite{faasse2020public}. Similarly, another study indicated that women generally exhibit greater levels of perceived risk and fear compared to men \cite{yildirim2021impacts}.

Our research expands upon prior studies on movement rhythms during the pandemic by utilizing a high-resolution, objective method for measuring movement rhythms, improving upon survey-based approaches. We suggest a straightforward approach to extract movement rhythms from fitness trackers, complementing existing techniques such as mapping smartphone locations \cite{gao2020mapping} and geo-located Twitter \cite{hawelka2014geo}. Additionally, our dataset allows for capturing movement rhythms over an extended time, enhancing the reliability of the proposed method.

\subsection*{Limitations}

Our work is not without limitations. Firstly, to measure step counts we use Polar Ignite devices which are consumer-grade fitness trackers. The calculation of step counts is based on Polar's proprietary algorithms and we do not have access to the raw and unprocessed data. However, as our work primarily relies on relative step counts throughout the day, rather than absolute numbers, we do not expect to have a large bias due to this. In the future, studies with research-grade monitors could provide more certainty in the absolute values of step counts when studying movement patterns. The second limitation is the geographic constraint. Between countries, physical activity level is greatly affected by the stringency of social distancing policies. In countries with stricter restriction, access to public spaces (parks, fitness centers, supermarkets, etc.) were denied at times. Moreover, residents in many countries were only allowed to go out for ``essential activities'' every two to three days \cite{celliniChangesSleepPattern2020, heChangesBodyWeight2021} during certain time periods. Thus, lockdowns might result in more consistent daily rhythms of movement (even if they lead to less amount of daily movement), since the residents are confined in their homes most of the time. Third, our cohorts mainly consist of Finland-based participants working at a higher institution. This limitation hinders detailed comparison of movement rhythms among people from different places and our results cannot be generalized to other societies, workplaces, or education levels. In the future, similar mixed-method data collection studies with participants from a wide range of sociodemographic backgrounds can use our methodology to quantify daily movement rhythms and their consistency over time and investigate the effect of different personal, work-place related, or societal level changes in life circumstances on those rhythms. 

\backmatter

\bmhead{Acknowledgments}

We acknowledge the computational resources provided by the Aalto Science-IT project. We thank Mikko Kivel\"a for providing thorough and helpful comments on our manuscript. We also thank Arsi Ik\"aheimonen and Yunhao Yuan for providing feedback to us.

\newpage
\bibliography{sn-bibliography}

\newpage
\begin{appendices}

\newpage
\section{Mixed model: Controlling for demographic heterogeneity}\label{secA2}

\subsection*{Model 2: Weekend movement consistency}

\begin{table}[!h] 
 \centering 
 \sffamily 
 \footnotesize 
 \begin{tabular}{lrr@{}l} 
 \toprule 
 
 & Est.      & 95\% CI  &   \\ \midrule
 (Intercept)                     & 0.22 &	-0.08 – 0.53 & \\ 
 Role (service staff)            & -0.08 &	-0.36 – 0.21 & \\ 
 Gender (male)                   & -0.16 &	-0.45 – 0.15 & \\ 
 Live alone (yes)                & \textbf{-0.33} &	\textbf{-0.61 – -0.05} & *\\ 
 Have children (yes)             & -0.10 &	-0.40 – 0.19 & \\ 
  
 Origin (migrant)                & -0.19	& -0.50 – 0.10	 & \\ 
 Age                             & 0.01 & -0.12 – 0.14    &   \\ 
 Have children (yes) x Gender (male) & 0.22 &	-0.26 – 0.70  &  \\ 
 \bottomrule 
 \end{tabular} 
 \caption{Results for socio-demographic variables predicting short and long-term workday consistency. Asterisks denote the significance of the results. *$p<0.05$, **$p<0.01$, ***$p<0.001$.} 
 \label{table:weekday-long-consistency} 
 \end{table}

\subsection*{Model 3: On-site work attendance}

\begin{table}[!h] 
 \centering 
 \sffamily 
 \footnotesize 
 \begin{tabular}{lrr@{}l} 
 \toprule 
 
 & Est.      & 95\% CI  &   \\ \midrule
 (Intercept)                     & \textbf{0.30} &	\textbf{0.22 – 0.37} & *** \\ 
 Long-term workday consistency       & \textbf{0.07} & \textbf{0.04 – 0.09} & *** \\
 Role (service staff)            & \textbf{-0.12} &	\textbf{-0.17 – -0.07} & *** \\ 
 Gender (male)                   & \textbf{0.12}  &	\textbf{0.06 – 0.17} & ** \\ 
 Live alone (yes)                & \textbf{0.08} &	\textbf{0.03 - 0.13} & ***\\ 
 Have children (yes)             & -0.10 &	-0.06 – 0.04 & \\ 
 Origin (migrant)                & \textbf{0.13}	& \textbf{0.08 – 0.18}	 & *** \\ 
 Age                             & \textbf{0.02} & \textbf{0.00– 0.04}    & *   \\ 
 Have children (yes) x Gender (male) & -0.01 &	-0.10 – 0.07  &  \\ 
 \midrule
 \textbf{Random effects} \\
 $\sigma^2$ & 0.08\\
 $\tau_{00_{month}}$ & 0.01\\
 ICC & 0.08\\
 $N _{month}$ & 11\\
 Observations & 912\\
 Marginal $R^2$ / Conditional $R^2$ & 0.151 / 0.223\\
 \bottomrule 
 \end{tabular} 
 \caption{Results for socio-demographic variables and long-term workday consistency predicting on-site work attendance rate. Asterisks denote the significance of the results. *$p<0.05$, **$p<0.01$, ***$p<0.001$.} 
 \label{table:onsite-work-attendance} 
 \end{table}

\newpage

\section{Routine-related questions}\label{sec_SI_Q}
\begin{table}[htp]
\sffamily

\begin{tabular}{@{}ll@{}}
\toprule
Routine              & Question                                                                                                                                                                                                 \\ \midrule
                     & \begin{tabular}[c]{@{}l@{}}During the pandemic; what percentage of your working time \\ did you spend on-site? \textbf{(b)} \end{tabular}                                                                           \\
                     &                                                                                                                                                                                                          \\
Workplace            & \begin{tabular}[c]{@{}l@{}}Prior to the pandemic; what percentage of your working time \\ did you spend on-site? \textbf{(b)} \end{tabular}                                                                              \\
                     &                                                                                                                                                                                                          \\
                     & \begin{tabular}[c]{@{}l@{}}If you worked in (part of) \textit{month} what percentage of your \\ working time  did you spend on-site ? \\ (do not answer if you did not work at all in \textit{month}) \textbf{(m)} \end{tabular}     \\ \midrule
                     & \begin{tabular}[c]{@{}l@{}}Prior to the pandemic; how many hours per week on average \\ did you walk? \textbf{(b)} \end{tabular}                                                                                       \\
                     &                                                                                                                                                                                                          \\
Walking              & \begin{tabular}[c]{@{}l@{}}During the pandemic; how many hours per week on average \\ have you walked? \textbf{(b)} \end{tabular}                                                                                      \\
                     &                                                                                                                                                                                                          \\
                     & \begin{tabular}[c]{@{}l@{}}In the late stages of the pandemic; how many hours per week \\ on average have you walked? \textbf{(e)} \end{tabular}                                                                       \\ \midrule
                     & \begin{tabular}[c]{@{}l@{}}Prior to the pandemic; how many hours per week on average \\ did you engage in non-walking exercise \\ (e.g., riding a bike; weightlifting; etc)? \textbf{(b)} \end{tabular}                 \\
                     &                                                                                                                                                                                                          \\
Non-walking exercise & \begin{tabular}[c]{@{}l@{}}During the pandemic; how many hours per week on average  \\ have you engaged in non-walking exercise \\ (e.g., riding a bike; weightlifting; etc)? \textbf{(b)} \end{tabular}                \\
                     &                                                                                                                                                                                                          \\
                     & \begin{tabular}[c]{@{}l@{}}In the late stage of the pandemic; how many hours per week \\ on average  have you engaged in non-walking exercise \\ (e.g., riding a bike; weightlifting; etc)? \textbf{(e)} \end{tabular} \\ \bottomrule
\end{tabular}
\caption{Questions about life routines. Questions marked with \textbf{(b)}, \textbf{(e)}, \textbf{(m)} are asked in the baseline, exit, and monthly surveys, respectively.}
\label{table:routine_questions}
\end{table}

\newpage
\section{Different segmenting strategies for computing distribution of steps}\label{secA4}

\begin{figure}[!hbt]
    \centering
    \includegraphics[width=\textwidth]{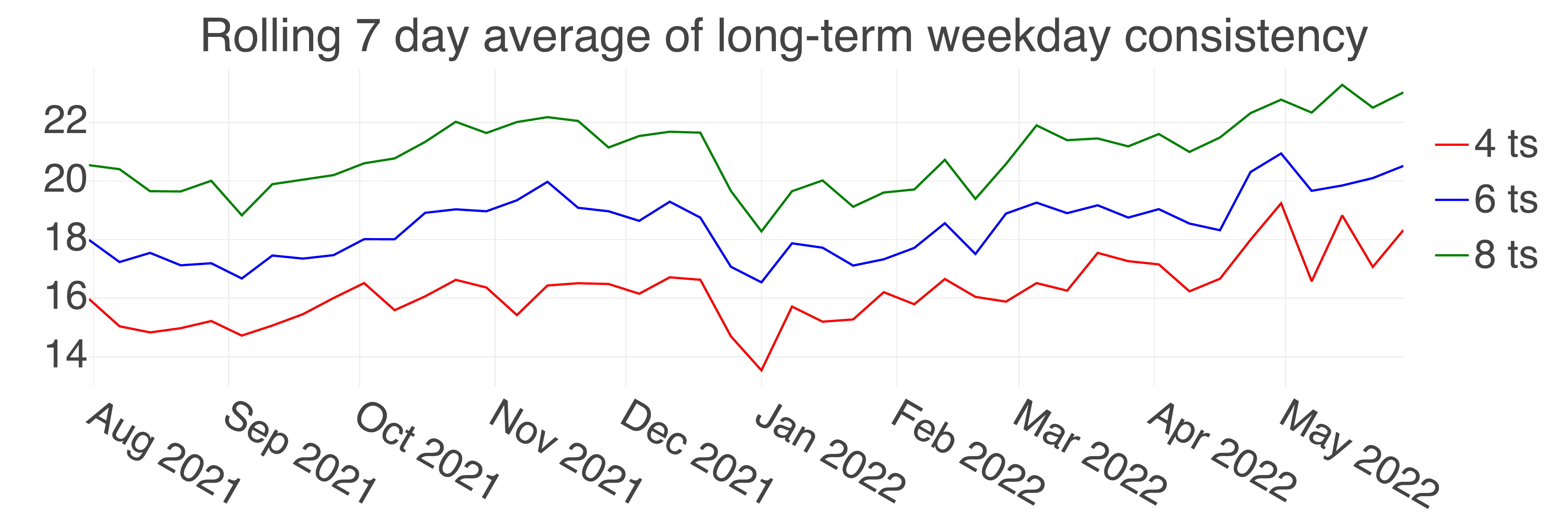}
    \caption{Long-term workday consistency with different segmenting strategies.}
    \label{fig:multiple_binned_consistency}
\end{figure}

In Figure \ref{fig:multiple_binned_consistency}, we demonstrate long-term workday consistency with different time segmenting strategies. The trend is represented as a 7-day rolling average. We can observe that the smaller the time segment (ts), the less drastic the change is. However, the general temporal trend holds regardless of the number of time segments.

\newpage
\section{Relationship between long-term movement rhythms consistency, onsite-work attendance, and stringency index}\label{secA5}

\begin{figure}[!hbt]
    \centering
    \includegraphics[width=\textwidth]{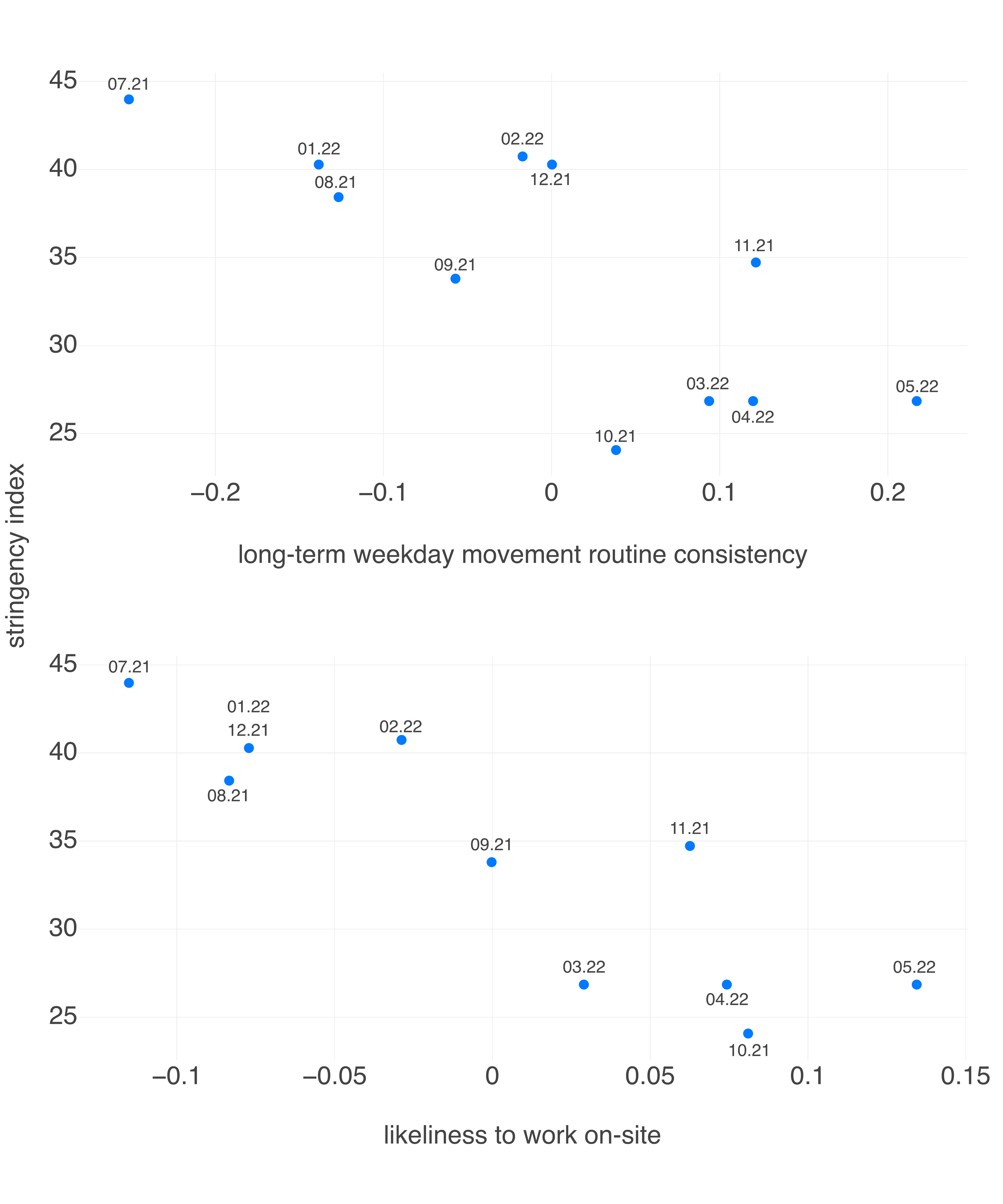}
    \caption{Relationship between stringency index and movement consistency and likeliness to work on-site. The random intercepts from Model 1b indicate the variability of the long-term workday movement consistency and the random intercepts from Model 3 indicate the likeliness to work on-site.} 
    \label{fig:si_vs_re}
\end{figure}

\end{appendices}

\end{document}